\documentclass{elsart}
\usepackage[dvips]{epsfig}
\usepackage{amsmath}
\usepackage{subfigure}

\begin{document}

\begin{frontmatter}


\title{Photon acceleration in vacuum}

\author[1,2]{J.T.\ Mendon\c{c}a\thanksref{tito}}
\author[2,3]{M.\ Marklund\thanksref{mattias}} 
\author[1,2,3,4,5]{P.K.\ Shukla\thanksref{padma}} and 
\author[2,3]{G.\ Brodin\thanksref{gert}}

\address[1]{CFP and CFIF, Instituto Superior T\'{e}cnico,1049-001 Lisboa, Portugal}
\address[2]{Centre for Fundamental Physics, Rutherford Appleton Laboratory,
  Chilton, Didcot, Oxon OX11 OQX, U.K.}
\address[3]{Nonlinear Physics Centre, Department of Physics, Ume{\aa} University, SE-90187 Ume{\aa}, Sweden}
\address[4]{Institut f\"ur Theoretische Physik IV and Centre for Plasma Science and Astrophysics,
Fakult\"at f\"ur Physik und Astronomie, Ruhr-Universit\"at Bochum, D--44780 Bochum, Germany}
\address[5]{Department of Physics, University of Strathclyde, Glasgow, Scotland, G4 ONG, UK}

\thanks[tito]{titomend@ist.utl.pt}
\thanks[mattias]{mattias.marklund@physics.umu.se}
\thanks[padma]{ps@tp4.rub.de}
\thanks[gert]{gert.brodin@physics.umu.se}

\date{\today}

\begin{abstract}

A new process associated with the nonlinear optical properties of the electromagnetic vacuum, 
as predicted by quantum electrodynamics, is described. This can be called photon acceleration 
in vacuum, and corresponds to the frequency shift that takes place when a given test photon 
interacts with an intense beam of  background radiation. \\

\end{abstract}

\end{frontmatter}


\section{Introduction}

It is known for a very long time that, in quantum electrodynamics (QED), photon-photon interactions 
in vacuum are possible due to the creation of virtual electron-positron pairs \cite{heisenberg,schwinger}. 
This leads to the appearance of nonlinear corrections of the photon dispersion relation in vacuum. 
In other words, the vacuum becomes a nonlinear medium. However, such corrections are extremely weak, 
and for this reason the systematic study of the vacuum nonlinearity only started recently. In particular, 
proposals for the experimental observation of such nonlinear corrections have been published. They include 
photon splitting \cite{birula,adler}, second harmonic generation \cite{ding}, self-focusing \cite{segev}, and 
nonlinear wave mixing in microwave cavities \cite{brodin01}. Also, attention has been paid to collective 
photon phenomena \cite{marklund-shukla}, such as the electromagnetic wave collapse \cite{mattias03,shukla04a} 
as well as the formation of photon bullets \cite{mattias04} and light wedges \cite{shukla04b}.

In this paper,  we explore another process associated with the collective photon interactions, related 
to  possible frequency shift of test photons immersed in a modulated radiation background. This new process 
could be called photon acceleration in vacuum, because of its obvious analogies with the well documented 
photon acceleration processes that can occur in a plasma or in an optical medium \cite{mendbook}. 
Using the Heisenberg--Euler Lagrangian density, the appropriate 
dispersion relation for a photon state in vacuum in the presence of a generic radiation background can be found. We will 
consider the case where the background is a nearly unidirectional photon beam. The photon ray equations pertinent for this case is studied in detail, and the 
resulting frequency shift is determined. An order of magnitude estimate for the expected frequency 
shift is presented, based on analytical and numerical calculations.

\section{Dispersion relation}

The nonlinear QED effects associated with the creation of virtual electron-positron pairs in vacuum can 
be described by the Heisenberg--Euler Lagrangian density, which can be written as the usual classical 
electromagnetic Lagrangian density $\mathcal{ L}_0$ plus a nonlinear quantum correction $\delta \mathcal{ L}$ 
in the form \cite{zuber}

\begin{equation}
\mathcal{ L} = \mathcal{ L}_0 + \delta \mathcal{ L} = \epsilon_0 \mathcal{ F} + \zeta (4 \mathcal{ F}^2 + 7 \mathcal{ G}^2), 
\label{eq:2.1} \end{equation}
where the quantities $\mathcal{ F}$ and $\mathcal{ G}$ are determined from
$ \mathcal{ F} \equiv \mathcal{ L}_0 / \epsilon_0 = \frac{1}{2} (E^2 - c^2 B^2)$ and 
$\mathcal{ G} = c (\vec{E} \cdot \vec{B})$. Here $\vec{E}$ and $\vec{B}$ are the electric and magnetic fields, respectively.  
The nonlinear parameter appearing in (\ref{eq:2.1}) is $ \zeta = 
2\alpha^2 \epsilon_0^2 \hbar^3/45 m_e^4 c^5$,
where $\alpha = e^2 / 2 \epsilon_0 h c \approx 1 / 137$ is the fine structure constant.
The QED corrections in the Lagrangian (\ref{eq:2.1}) are valid in the weak field, low frequency approximation, i.e. we require $|E| \ll E_{\mathrm{crit}}$ and $\omega \ll \omega_e = m_ec^2/\hbar$, where $E_{\mathrm{crit}} \approx 10^{16}\,\mathrm{V/cm}$ is the Schwinger critical field, $\omega_e$ is the Compton frequency, and $\omega$ is the typical frequency with which the electromagnetic field changes. These approximations ensures that there is no appreciable pair creation due to multi-photon effects (as it will be exponentially suppressed for low field strengths) and that there are no single photons able to generate pairs from the vacuum. However, it is worth pointing out that we do \emph{not} require the fields to be constant in time or space, only slowly varying with respect to the Compton frequency \cite{schwinger}. As this is the case for almost all relevant fields, the applicability of the Lagrangian (\ref{eq:2.1}) is guaranteed for a wide variety of field configurations. 
The resulting Maxwell's equations in vacuum take the usual form, if we use the following definitions
$\vec{D} = \epsilon_0 \vec{E} + \vec{P}$ and $\vec{B} = \mu_0 (\vec{H} + \vec{M})$,
where the polarization $\vec{P}$ and magnetization $\vec{M}$ result from the nonlinear QED corrections 
included in (\ref{eq:2.1}), and are given by

\begin{equation}
\vec{P} = 2 \zeta (4 \mathcal{ F} \vec{E} + 4 c \mathcal{ G} \vec{B} )
\quad , \quad
\vec{M} = - 2 c^2 \zeta (4 \mathcal{ F} \vec{B} + 7 \mathcal{ G} \vec{E} / c).
\label{eq:2.5} \end{equation}
Starting from Maxwell's equations in vacuum, we can then establish the equation of propagation for 
the electric field $\vec{E}$ in the form

\begin{equation}
\left( \nabla^2 - c^{-2} {\partial_t^2} \right) \vec{E} 
= \mu_0 \left[ \partial_t^2 \vec{P} + c^2 \nabla ( \nabla \cdot \vec{P} ) 
+ \partial_t (\nabla \times \vec{M} ) \right],
\label{eq:2.6} \end{equation}
and a similar equation for the magnetic field. If we consider a given photon state, with the frequency $\omega$ 
and the wavevector $\vec{k}$, we can derive from these wave equations the following nonlinear dispersion 
relation  \cite{birula,brodin01}

\begin{equation}
\omega = k c \left( 1 - \tfrac{1}{2}\lambda | Q |^2 \right),
\label{eq:2.7} \end{equation}
where $Q$ represents the background electromagnetic field. We have  
$| Q |^2 = \epsilon_0 \left[ E^2 + c^2 B^2 - (\vec{n} \cdot \vec{E} )^2 - c^2 (\vec{n} \cdot \vec{B} )^2 
- 2 c \vec{n} \cdot (\vec{E} \cdot \vec{B} ) \right]$
and $\lambda = \lambda_{\pm}$, with $\lambda_{+} = 14 \zeta$ and $\lambda_{-} = 8 \zeta$ for the two 
independent photon polarization states. We have also used the unit propagation vector $\vec{n} = \vec{k} / k$. We note that Eq.\ (\ref{eq:2.7}), to first order in the QED corrections, can be rearranged to give the wavenumber as a function of frequency (both forms being equivalent). However, the current form will be used in Sec.\ III (see Eq.\ (9) and the discussion following it) as it is the canonical choice using the geometric optics approximation. 

It may also be argued that derivative corrections to the weak field Heisenberg-Euler Lagrangian (\ref{eq:2.1}) could be of importance in certain applications, especially for high-frequency field configurations where the local field approximation may be questionable. However, the derivative correction to the weak field Lagrangian (\ref{eq:2.1}) was given in Ref.\ \cite{mamaev}, taking the form
\begin{equation}
  \delta\mathcal{L}_D = \frac{2\epsilon_0c^2\alpha}{15\omega_e^2}\left[
    (\partial_aF^{ab})(\partial_cF^c\!_b) - F_{ab}\partial^2F^{ab}
  \right] ,
\end{equation}
where $F_{ab}$ is the field strength tensor and we have used the summation convention over the four-indices $a,b,c$, 
and this correction gives rise to terms in the dispersion relation proportional to the parameter 
$\alpha^2(\omega/\omega_e)^{2}(|E|/E_{\mathrm{crit}})^{2}$ \cite{Rozanov}. Thus, for frequencies much smaller than the Compton frequency, appropriate for the field configurations considered below, these derivative correction are smaller by a factor $\alpha(\omega/\omega_e)^2 \ll 1$ than the QED corrections in the Lagrangian (\ref{eq:2.1}), and they may therefore at this stage be safely neglected in comparison to the correction in (\ref{eq:2.1}) (that are proportional to $\alpha(|E|/E_{\mathrm{crit}})^{2}$). However, on longer time-scales such corrections could in principle yield interesting dispersive effects on the interaction among photons \cite{mattias04,shukla04b}. For a thorough discussion of the Euler-Heisenberg Lagrangian and its validity limits, see e.g. Ref. \cite{D-and-G}

Let us now consider the simplest case where the background radiation field is associated with a single 
photon state with the frequency $\omega'$ and the wavevector $\vec{k}'$. Here the dispersion relation (\ref{eq:2.7}) 
takes the form

\begin{equation}
\omega = k c \left[ 1 - \tfrac{1}{2} \epsilon_0\lambda f (\vec{k}, \vec{k}') | E (\vec{k}') |^2 \right],
\label{eq:2.9} \end{equation}
where $ \vec{E} (\vec{k}')$ is the electric field of the background radiation, and $f (\vec{k}, \vec{k}')$ 
is a geometric factor determined by

\begin{equation}
f (\vec{k}, \vec{k}') = \left\{ 2  - (\vec{n} \cdot \vec{e}' )^2 
-  [ \vec{n} \cdot (\vec{n}' \times \vec{e}' ) ]^2 - 2  (\vec{n} \cdot \vec{n}' )  ] \right\}
\label{eq:2.10} \end{equation}
involving the unit vector $\vec{n}$ defined above, as well as $\vec{n}' = \vec{k}' / k'$ and 
$\vec{e}' =  \vec{E} (\vec{k}') /  E (\vec{k}')$. It is obvious that, for $\vec{n} = \vec{n}'$, we 
have $f (\vec{k}, \vec{k}')  = 0$, which means that there is no coupling between photons belonging to the 
same photon state. In other words, the nonlinear self-interaction of a photon with itself is strictly 
forbidden (to all orders in the amplitude, see Ref.\ \cite{schwinger}). 
Only different states of the radiation background will influence the dispersion of a photon 
in vacuum, through the virtual electron--positron pair fluctuations. A test photon moving in a background field will generically have a non-vanishing  geometric factor (\ref{eq:2.10}). In particular, a background field  consisting of a plane wave will  result in a  non-vanishing geometric factor  unless the  test photons are propagating parallel to the wave vector of the background field. We emphasize that the finite value of the geometric factor (10) is a direct consequence of non-vanishing values of the relativistic invariants F and G when considering the total electromagnetic field, i.e. incorporating both the background fields and the test photon field in the calculation scheme.

We can easily generalize the dispersion relation (\ref{eq:2.9}) for the case of an arbitrary broadband 
radiation spectrum, which can be slowly modulated in space and time over length and time scales much 
longer than the periods and the wavelengths of the considered radiation states. The resulting expression is 

\begin{equation}
\omega (\vec{r}, \vec{k}, t) = k c \left[ 1 - \frac{\lambda^*}{2} \int f (\vec{k}, \vec{k}') k' 
N (\vec{r}, \vec{k}', t) \frac{d \vec{k}'}{(2 \pi)^3} \right],
\label{eq:2.11} \end{equation}
where we have introduced the photon occupation number, or number of photons per field mode, as determined by

\begin{equation}
N (\vec{k}') = \frac{\epsilon_0 c}{4 \hbar k'} | E (\vec{k}') |^2,
\label{eq:2.12} \end{equation}
and a new coupling factor determined by $\lambda^* = 4 \hbar \lambda / c$.

\section{Photon dynamics}

In the geometric optics approximation, the dynamics of photons can be described by the ray equations, 
which can be stated in the following canonical form \cite{Whitham}

\begin{equation}
\frac{d \vec{r}}{d t} = \frac{\partial \omega}{\partial \vec{k}} \quad , \quad
\frac{d \vec{k}}{d t} = - \frac{\partial \omega}{\partial \vec{r}},
\label{eq:3.1} \end{equation}
where the Hamiltonian $\omega$ is determined by Eq. (\ref{eq:2.11}). From (\ref{eq:3.1}) follows that $d\omega/dt = \partial\omega/\partial t$, and in accordance with the general principles of geometrical optics, the ray equations imply that a time-dependent medium is needed for photon acceleration (frequency conversion) to occur, i.e. for a pure space-dependence of the Q-factor in (4) the photon frequency is conserved. 

Let us consider the important 
particular case where the background radiation is dominated by a single photon beam propagating in 
a direction $Oz$. This means that we can make $\vec{k}' = k' \vec{e}_z$. If the probe photon described 
by these equations of motion propagates at a given angle $\theta$ with respect to the $z$-axis, we 
have $(\vec{n} \cdot \vec{n}' ) = \cos \theta$ and $(\vec{n} \cdot \vec{e}' ) = \sin \theta \cos \psi$, 
where $\psi$ is the second angle necessary to define the direction of $\vec{n}$ with respect to the 
background electric field. From (\ref{eq:2.10}), we obtain the geometric factor
$f (\vec{k}, \vec{k}') \equiv f (\theta) = [ 2 (1 - \cos \theta) - \sin^2 \theta ]$.
The photon dispersion relation  then becomes

\begin{equation}
\omega (\vec{r}, \vec{k}, t) = k c \left[ 1 - \tfrac{1}{2} \lambda^*f (\theta) I (\vec{r}, t) \right],
\label{eq:3.3} \end{equation}
where the angle $\theta$ can vary in space and time, according to the photon dynamics described by 
Eq. (\ref{eq:3.1}), and the background field intensity $I (\vec{r}; t)$ is determined by the integral

\begin{equation}
I (\vec{r}, t) = \int k' N (\vec{r}, \vec{k}', t) \frac{d \vec{k}'}{(2 \pi)^3}. 
\label{eq:3.4} \end{equation}

Let us assume that the beam is modulated in intensity along the propagation direction, and that we can write 
$I (z - u t, \vec{r}_\perp)$. For a very large beam waist, we can neglect the dependence over the transverse 
direction and assume that $u = c$. From the equations of motion (\ref{eq:3.1}) we can then conclude that 
the transverse photon dynamics is an invariant

\begin{equation}
\frac{d \vec{k}_\perp}{d t} = - \frac{\partial \omega}{\partial \vec{r}_\perp} = 0
\quad , \quad k_\perp = k \sin \theta = \text{constant}.
\label{eq:3.5} \end{equation}
Even if this is not exact, a small variation in $k_\perp$ would only result in a slight diffraction effect, 
with no significant contribution to the frequency shift that will be examined below. In the general case, 
however, both $k$ and $\theta$ will evolve in space and time, as determined by the parallel equations of motion

\begin{equation}
\frac{d z}{d t} = \frac{\partial \omega}{\partial k_z} 
= \frac{k_z c}{k} \left[ 1 - \frac{\lambda^*}{2} f (\theta) I (z - c t) \right],
\label{eq:3.6} \end{equation}
and
\begin{equation}
\frac{d k_z}{d t} = - \frac{\partial \omega}{\partial z} 
= k c \frac{\lambda^*}{2} f (\theta) \frac{\partial}{\partial z} I (z - c t). 
\label{eq:3.6b} \end{equation}

The parallel photon motion can be more appropriately described by using a new variable $\eta = z - c t$. 
This can be formally done by making a canonical transformation from $(z, k_z)$ into a new pair of 
variables $(\eta, p)$, through the generating function $F (z, p, t) = (z - c t) p$. Hence, we have

\begin{equation}
k_z = \frac{\partial F}{\partial z} = p \quad , \quad
\eta = \frac{\partial F}{\partial p} = z - c t.
\label{eq:3.7} \end{equation}
The resulting new Hamiltonian will then be defined by

\begin{equation}
\omega' (\eta, p) = \omega (\eta, p) + \frac{\partial F}{\partial t},
\label{eq:3.8} \end{equation}
or, in explicit form, by
\begin{equation}
\omega' (\eta, p) = c \sqrt{k_\perp^2 + p^2}  \left[ 1 - \frac{\lambda^*}{2} 
f (\theta) I (\eta ) \right] - c p,
\label{eq:3.8b} \end{equation}
where $k_\perp$ is a constant, as stated by equation (\ref{eq:3.5}). The canonical equations for 
the parallel photon motion become

\begin{equation}
\frac{d \eta}{d t} = \frac{\partial \omega'}{\partial p} 
= - c + \frac{p c}{k} \left[ 1 - \frac{\lambda^*}{2} f (\theta) I (\eta) \right], 
\label{eq:3.9} \end{equation}
and 
\begin{equation}
\frac{d p}{d t} = - \frac{\partial \omega'}{\partial \eta} 
= k c \frac{\lambda^*}{2} f (\theta) I' (\eta). 
\label{eq:3.9b} \end{equation}

In order to understand the implications of these equations, we will assume that the beam intensity 
is modulated over a distance $\Delta$, as described explicitly by the function

\begin{equation}
I (\eta ) \equiv I_0 g (\eta ) = \frac{I_0}{2} \left[ 1 \pm \tanh (\eta / \Delta) \right], 
\label{eq:3.10} \end{equation}
where the signs $+$ and $-$ can describe the rear and front of the beam , respectively.

Before describing any explicit solution, it is however important to notice that the new Hamiltonian 
$\omega'$ is an invariant, which allows us to establish a relation between the values of the photon 
frequency at two distinct positions of its trajectory. Let us then consider two different positions 
along the same photon trajectory, viz. $\eta = \eta_1$ and $\eta = \eta_2$. Because of the invariance 
of $\omega'$, we can write $\omega' = k_1 c ( 1 - \delta n_1) - c p_1 = k_2 c (1 - \delta n_2) - c p_2$,
where we have used $k_i = k (\eta_i)$ and $p_i = p (\eta_i)$, for $i = 0, 1$, and defined the 
nonlinear refractive index perturbations $\delta n_i = \tfrac{1}{2}\lambda^* f (\theta_i ) I (\eta_i )$.

Notice that, far ahead of (or far back of) the background beam we have $\delta n \approx 0$, and deep 
inside the beam we attain the maximum perturbation $\delta n_{\mathrm{max}} \approx (\lambda^* / 2) f (\theta) I_0$. 
This expression can be explicitly written in terms of the photon frequencies $\omega_i$, noticing that they 
are related to the wavenumbers $k_i$ through the dispersion relation (\ref{eq:3.3}), which can be rewritten as

\begin{equation}
\omega_i = k_i c ( 1 - \delta n_i) \approx \frac{k_i c}{1 + \delta n_i}.  
\label{eq:3.13} \end{equation}
This leads to the following relation between the initial value of the frequency $\omega_0 \equiv \omega (t = 0)$, 
and a subsequent value $\omega_1 = \omega (t)$

\begin{equation}
\omega (t) = \omega_0 \; \frac{1 - \delta n (t)}{(1 - \delta n_0)} \; 
\frac{(1 - \cos \theta_0 - \delta n_0)}{1 - \cos \theta (t) - \delta n (t)}.
\label{eq:3.14} \end{equation}
This expression will be very useful to calculate the frequency shift of a test photon due to its interaction 
with the radiation background. Such a frequency shift can be associated with a process of photon acceleration. 
Notice that the group velocity of the photon as determined by the nonlinear dispersion relation, 
and given by \cite{birula} $\vec{v} = \vec{n} c [ 1 - \delta n (t) ]$, 
is independent of the photon frequency, but changes with time due to the change in the angle between 
the photon and the background beam. In order to determine how this process evolves with time, we have to 
go back to the parallel photon dynamical equations. It is useful to rewrite them using a new time variable, 
defined by $\tau = c t \; \lambda^* I_0 / 2$, which describes the photon evolution on a very slow time scale. 
Introducing the dimensionless variable $\zeta = p / k_\perp$, we have then

\begin{equation}
\frac{d \eta}{d \tau} = - \left[ 2 \frac{(1 - X)}{\lambda^* I_0} + X (1 - X)^2 g (\eta) \right],
\label{eq:3.16} \end{equation}
and 
\begin{equation}
\frac{d \zeta}{d \tau} =  (1 - X)^2 \sqrt{1 + \zeta^2}  g' (\eta),
\label{eq:3.16b} \end{equation}
where $X \equiv \cos \theta =\zeta / \sqrt{1+ \zeta^2}$. For a plausible situation, even for very intense 
background radiation intensity, such that $\lambda^* I_0 \ll (1 - X)$, we can use the approximate solution

\begin{equation}
\eta (\tau) \approx - \frac{2 [1 - X (0)]}{\lambda^* I_0} \tau \equiv - u_0 \tau,
\label{eq:3.17} \end{equation}
and, using Eq. (\ref{eq:3.10}), we can write

\begin{equation}
\frac{d \zeta}{d \tau} \approx \pm [1 - X (\zeta)]^2 \frac{1}{\Delta}  \sqrt{1 + \zeta^2} 
\; sech^2 (- u_0 \tau / \Delta).
\label{eq:3.18} \end{equation}

This approximate description of the photon acceleration process occurring in vacuum, on a very slow time-scale 
defined by $\tau$, is completed by Eq. (\ref{eq:3.14}). It is nearly equivalent to

\begin{equation}
Y (\tau) \equiv \frac{\omega ( \tau)}{\omega (0)} \approx \frac{1 - X (0)}{1 - X (\tau)}.
\label{eq:3.18b} 
\end{equation}

\begin{figure*}
\subfigure[]{\epsfig{file=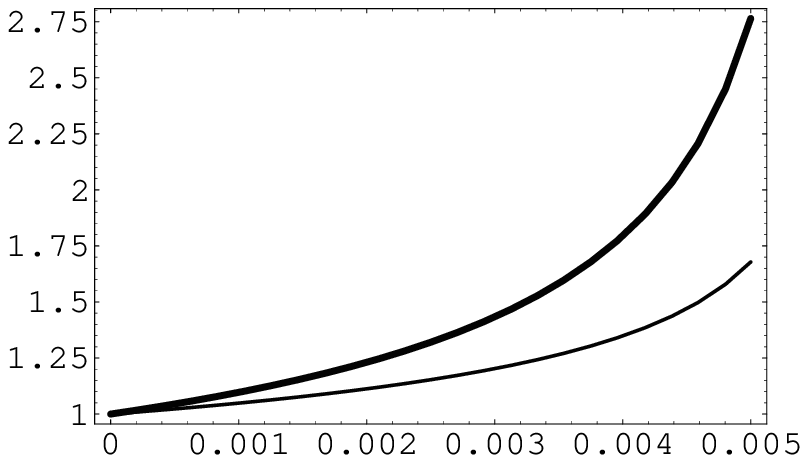,width=.48\textwidth,clip=}}
\subfigure[]{\epsfig{file=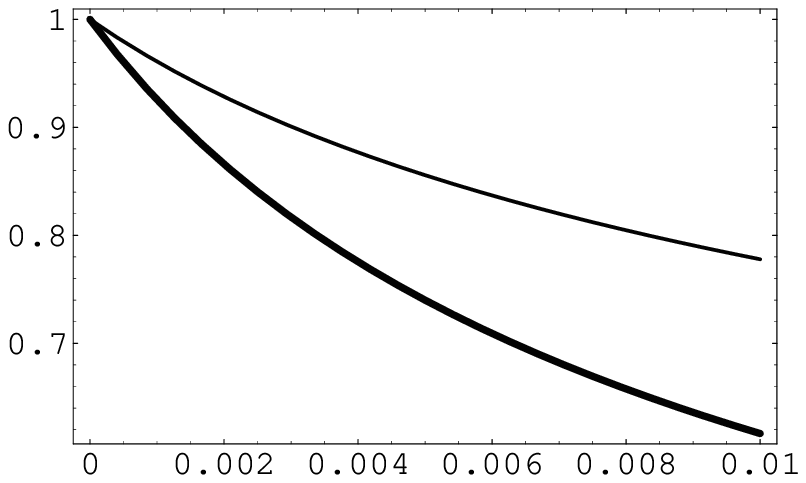,width=.48\textwidth,clip=}}
\caption{(a) Photon acceleration process at the rear of an intense beam in vacuum: Evolution of the relative 
parallel wavenumber $\zeta (\tau ) / \zeta (0)$, and the relative frequency shift $Y (\tau) $ (in bold), as a function of 
$\tau / \Delta$, for $u_0 = 1$,  for an initial value $\zeta (0) = 5$. (b) Photon deceleration process at the front of an intense beam in vacuum: Evolution of the 
relative parallel wavenumber $\zeta (\tau) / \zeta (0)$, and the relative frequency shift $Y (\tau)$ (in bold), as a function 
of $\tau / \Delta$, for $u_0 = 1$,  for an initial value $\zeta(0) = 5$.}
\label{fig:fig1}
\end{figure*}

The results are illustrated in Fig.\ 1a , where the values of $\zeta (\tau)$ and $Y (\tau)$ are represented 
for photons with initial values $\zeta (0) = 5$. Here the photon is interacting with the rear 
of the background beam, and we can see that a noticeable up-shift of the photon frequency can take place. 
This is an interesting result, given the extremely small value of the nonlinear QED term responsible for 
the interaction. The physical meaning is, however, very clear: the very weak nonlinear force acting on 
the test photon and due to the gradient of the beam intensity (represented by the derivative $g' (\eta)$) 
acts on the photon over extremely large distances, because it travels with  a parallel velocity nearly 
equal to the light speed $c$, thus leading to a non-negligible frequency up-shift. 
Notice that the approximate solution (\ref{eq:3.18}) gives an overestimate of the relative velocity $u_0$, 
and therefore underestimates the up-shift. The frequency shifts tend to increase for lower values of $\zeta (0)$.


The effect of a frequency down-shift can occur at the front of the beam, as illustrated in Fig.\ 1b. We see that, 
in this case, the photon frequency deceases, and can eventually tend to zero, for lower initial values of $\zeta$.

\section{Conclusions}

We have described here a new process associated with the nonlinear optical properties of the electromagnetic 
vacuum, as predicted by quantum electrodynamics. This can be called photon acceleration in vacuum, and can 
take place when a given test photon interacts with a modulated background radiation. This process is similar 
to that already observed in laboratory experiments, using relativistic ionization fronts \cite{dias}. Here, 
however, there is no material supporting the interaction, except for the QED nonlinearities of vacuum. 
The ionization front is replaced by a trail of virtual electron--positron pairs created by the intense background 
radiation in vacuum. In the present case we have a kind of virtual ionization front, which is responsible for 
the photon frequency shift.

The case when the background is made of a single intense beam, propagating along a given direction, was 
considered in detail. It was shown that, over long interaction distances, the direction of propagation 
and the frequency of test photon can be significantly changed. A frequency up-shift occurs at the rear 
of the beam, and a down-shift can take place at the front. A similar process can occur when a photon 
interacts with an acoustic perturbation of the background, and will be the object of a forthcoming publication. 
In this case, a stronger effect is to be expected, because the geometric factor $f (\theta)$ will be larger 
than in the present case of nearly parallel photon-photon interaction.

In the present work only individual photon dynamics was considered. But the test pulse can belong to the 
radiation background itself, which means that the process described here will lead to spectral changes of 
the radiation spectrum in vacuum. This spectral evolution can be conveniently treated with the help of 
photon kinetics \cite{mendbook}, to be analyzed in the future. 


\end{document}